\documentclass[aps,10pt,letterpaper,twocolumn,english,showpacs,preprintnumbers,amsmath,amssymb,nofootinbib,prl]{revtex4-1}
\usepackage{amsmath}
\usepackage{graphicx}
\usepackage{setspace}
\usepackage{amssymb}
\usepackage{dcolumn}
\usepackage{mathrsfs}
\usepackage{bm}
\usepackage{longtable}

\begin{document}

\preprint{}

\title{Temporal Quantum-State Tomography of Narrowband Biphotons}

\author{Peng Chen}
\author{Chi Shu}
\author{Xianxin Guo}
\author{M. M. T. Loy}
\author{Shengwang Du} \email{dusw@ust.hk}
\affiliation{Department of Physics, The Hong Kong University of Science and Technology, Clear Water Bay, Kowloon, Hong Kong, China}

\date{\today}% It is always \today, today,
             %  but any date may be explicitly specified

\begin{abstract}
We describe and demonstrate a quantum state tomography for measuring the complex temporal waveform of narrowband biphotons. Through six sets of two-photon interference measurements projected in different polarization subspaces, we can construct the time-frequency entangled two-photon joint amplitude and phase functions in continuous-variable time domain. For the first time, we apply this technique to experimentally determine the temporal quantum states of narrowband biphotons generated from spontaneous four-wave mixing in cold atoms, and fully confirm the theoretical predictions.
\end{abstract}


%03.65.Wj	State reconstruction, quantum tomography
%03.67.Mn	Entanglement measures, witnesses, and other characterizations (see also 03.65.Ud Entanglement and quantum nonlocality; 42.50.Dv Quantum state engineering and measurements in quantum optics)
%42.50.Dv	Quantum state engineering and measurements (see also 03.65.Ud Entanglement and quantum nonlocality, e.g., EPR paradox, Bells inequalities, GHZ states, etc.)

\maketitle

%%%%%%%% introductory paragraph %%%%%%%%%%%%%%%%%%%%%%%%%%

Photons are described by their discrete polarization states and field amplitude distribution in continuous time-space domains. The density matrix of a polarization state can be reconstructed using well-developed quantum state tomography \cite{KwiatPRL1999, WhitePRA2001, AdamsonPRL2010}. To characterize the temporal modes of photons, one needs continuous-variable optical quantum-state tomography that is much more complicate than that for the discrete Hilbert space with finite dimensions. Homodyne detection has been proven an efficient probe to characterize photonic (unentangled) Fock and coherent states \cite{LvovskyPRL2001, OurjoumtsevPRL2006, MacRaePRL2012, LvovskyRMP2009, LauratPRL2013, Lvovsky2014}. However, most homodyne measurements for bipartite two-mode squeezed states have aimed only to verify their entanglements \cite{OuPRL1992, SchoriPRA2002, LauratJOB2005,VasilyevPRL2000, DAuriaPRL2009}, and a complete optical homodyne tomography for time-frequency entangled two-photon (amplitude and phase) temporal waveform has never been demonstrated \cite{LvovskyRMP2009}.

Developing narrowband biphoton source with high spectral brightness recently has attracted much attention because of the need for realizing efficient light-matter quantum interface \cite{PanPRL2008, ScholzPRL2009, FeketePRL2013}. Using spontaneous four-wave mixing (SFWM) with electromagnetically induced transparency (EIT) in cold atoms, sub-MHz biphoton generation with a coherence time up to microsecond has been demonstrated \cite{Subnatural, YanPRL2014, DuOptica2014}. Such a long coherence time of single photons allows manipulating their temporal waveform \cite{EOMSinglephoton, DuPRA2009, DuPRL2010Biphoton} as well as their interaction with atoms in time domain \cite{SinglePhotonPrecursor, DuOptExpress2012, DuPRL2012}. Owning to the time resolution of single photon counting modules (SPCM), the biphoton amplitude temporal profile can be directly measured as time-resolved coincidence counts. However, the photon coincidence counting does not tell the difference between a time-frequency entangled state and a temporal-probability mixed state because it measures only the joint probability distribution. Although additional evidences, such as violation of the Cauchy-Schwartz inequality \cite{ClauserPRD1974} and measurement of autocorrelation of heralded single photons \cite{GrangierEPL1986}, can indirectly confirm the biphoton nonclassical properties, a complete temporal tomography of narrowband biphoton waveform (amplitude + phase) has not been demonstrated. It is believed that the time-frequency entanglement of the paired photons in a continuous-wave (cw) SFWM is naturally endowed by energy conservation \cite{DuJOSAB2008}, but this claim has not been directly confirmed experimentally.

In this Letter, we propose and demonstrate a temporal quantum state tomography for SFWM narrowband biphotons. Using six sets of symmetrized time-resolved two-photon interference measurements in different polarization basis, we construct both the amplitude and phase functions of the biphoton joint temporal waveform.

\begin{figure}
\includegraphics[width=0.9\linewidth]{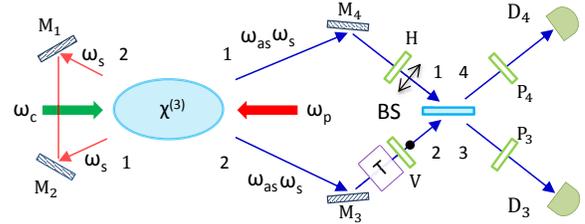}
\caption{\label{fig:Schematics} (color online). Schematics of temporal state tomography for narrowband biphotons generated from backward-wave spontaneous four-wave mixing.}
\end{figure}

Our modeled system is illustrated schematically in Fig.\ref{fig:Schematics}. With counter-propagating pump ($\omega_p$) and coupling ($\omega_c$) laser beams, phase-matched Stokes ($\omega_s$) and anti-Stokes ($\omega_{as}$) paired photons are produced in opposing directions. The single-spatial-mode biphoton temporal (complex) waveform output from the source is described as
\begin{eqnarray}
\Psi(t_s,t_{as})&=&\langle0|\hat{a}_{s}(t_{s})\hat{a}_{as}(t_{as})|\Psi_{s,as}\rangle \nonumber\\
&=&e^{-i\omega_{s0}t_s}e^{-i\omega_{as0}t_{as}}\psi(\tau), \label{eq:Waveform0}
\end{eqnarray}
where $\tau=t_{as}-t_s$, $\omega_{s0}$ and $\omega_{as0}$ are the central optical angular frequencies of Stokes and anti-Stokes photons, respectively. The relative biphoton waveform $\psi(\tau)$ can be rewritten as
\begin{eqnarray}
\psi(\tau)=A(\tau)e^{i\phi(\tau)}, \label{eq:RelativeWaveform}
\end{eqnarray}
where $A(\tau)$ is the amplitude and $\phi(\tau)$ is the phase. The amplitude function can be directly determined from the Glauber correlation function $G^{(2)}(\tau)=A^2(\tau)$ measured as two-photon coincidence. Due to the time ordering of the SFWM process and slow light effect on the anti-Stokes photons, $A(\tau)$ is nonzero only at $\tau>0$ \cite{DuJOSAB2008}. That is, the anti-Stokes photon is always generated after its paired Stokes photon. Our major task of the work is to determine the phase $\phi(\tau)$ by two-photon interference.

With the geometry shown in Fig.\ref{fig:Schematics}, phase matching allows generating photon pairs into two symmetric paths (1 and 2). We use two mirrors ($M_1$ and $M_2$) to counter-overlap the two Stokes modes. In this configuration, there are Stokes and anti-Stokes photons in each output port 1 or 2.  The photons are set to horizontally (H: $\leftrightarrow$) and vertically (V: $\updownarrow$) polarized at output ports 1 and 2 respectively. After a relative time delay of $T$ in path 2, the photons are then combined together at a 50\%:50\% beam splitter (BS), whose outputs 3 and 4 pass the polarization selectors (P$_3$ and P$_4$) and are detected by two SPCMs (D$_3$ and D$_4$). The combination of the BS and polarization selectors can induce two-photon interference by erasing the path and polarization information in a controllable degree of freedoms and is the key to reconstruct the phase information of the biphoton waveform.

The two-frequency-mode field operators at the source outputs can be written as $\hat{a}_{1\leftrightarrow}=\hat{a}_{1s\leftrightarrow}+\hat{a}_{1as\leftrightarrow}$ and $\hat{a}_{2\updownarrow}=\hat{a}_{2s\updownarrow}+\hat{a}_{2as\updownarrow}$. The BS has the transformation $\hat{a}_3=(\hat{a}_{1\leftrightarrow}+i\hat{a}_{2\updownarrow})/\sqrt{2}$ and $\hat{a}_4=(i\hat{a}_{1\leftrightarrow}+\hat{a}_{2\updownarrow})/\sqrt{2}$. The projection of the polarization selectors can be characterized as $\hat{a}_{Pm}=\hat{a}_{m\leftrightarrow}\cos\alpha_m+\hat{a}_{m\updownarrow}\sin\alpha_m e^{i\theta_m}$ ($m$=3, 4). The two-photon wave function at the two detectors can be derived as
\begin{widetext}
\begin{eqnarray}
\Psi_{34}(t_3,t_4)%&=&\langle0|-\cos\alpha_3\sin\alpha_4e^{i\theta_4}(\hat{a}_{1s\leftrightarrow}\hat{a}_{2as\updownarrow}\nonumber+\hat{a}_{1as\leftrightarrow}\hat{a}_{2s\updownarrow})+\sin\alpha_3\cos\alpha_4e^{i\theta_3}(\hat{a}_{2as\updownarrow}\hat{a}_{1s\leftrightarrow}+\hat{a}_{2s\updownarrow}\hat{a}_{1as\leftrightarrow})|\Psi\rangle \nonumber \\
%&=&-\cos\alpha_3\sin\alpha_4e^{i\theta_4}[e^{-i\omega_{s0}t_3}e^{-i\omega_{as0}(t_{4}-T)}\psi(t_{4}-T-t_3)+e^{-i\omega_{as0}t_{3}}e^{-i\omega_{s0}(t_4-T)}\psi(t_{3}-t_4+T)]\nonumber\\
%&+&\sin\alpha_3\cos\alpha_4e^{i\theta_3}[e^{-i\omega_{as0}(t_{3}-T)}e^{-i\omega_{s0}t_4}\psi(t_{3}-T-t_4)+e^{-i\omega_{s0}(t_3-T)}e^{-i\omega_{as0}t_{4}}\psi(t_{4}-t_3+T)]\nonumber\\
%&=&-\cos\alpha_3\sin\alpha_4e^{i\theta_4}[e^{-i\omega_{s0}t_3}e^{-i\omega_{as0}(t_{4}-T)}\psi(\tau-T)+e^{-i\omega_{as0}t_{3}}e^{-i\omega_{s0}(t_4-T)}\psi(-\tau+T)]\nonumber\\
%&+&\sin\alpha_3\cos\alpha_4e^{i\theta_3}[e^{-i\omega_{as0}(t_{3}-T)}e^{-i\omega_{s0}t_4}\psi(-\tau-T)+e^{-i\omega_{s0}(t_3-T)}e^{-i\omega_{as0}t_{4}}\psi(\tau+T)]\nonumber\\
&=&\frac{1}{2}\cos\alpha_3\sin\alpha_4e^{i\theta_4}e^{-i\omega_{s0}(t_3+t_4)}e^{i\omega_{s0}T}e^{-i\delta t_{3}}[e^{-i\delta(\tau-T)}\psi(\tau-T)+\psi(-\tau+T)]\nonumber\\
&-&\frac{1}{2}\sin\alpha_3\cos\alpha_4e^{i\theta_3}e^{-i\omega_{s0}(t_3+t_4)}e^{i\omega_{s0}T}e^{-i\delta t_{3}}[e^{i\delta T}\psi(-\tau-T)+e^{-i\delta \tau}\psi(\tau+T)].
\label{eq:Waveform34}
\end{eqnarray}
\end{widetext}
Here we define $\tau=t_4-t_3$ and $\delta=\omega_{as0}-\omega_{s0}$. Making use of Eq.(2) we obtain
\begin{widetext}
\begin{eqnarray}
\psi_{34}(T,\tau)&=&\Psi_{34}(t_3,t_4)e^{i\omega_{s0}(t_3+t_4)}e^{-i\omega_{s0}T}e^{i\delta t_{3}}=\frac{1}{2}\cos\alpha_3\sin\alpha_4e^{i\theta_4}[e^{-i\delta(\tau-T)}A(\tau-T)e^{i\phi(\tau-T)}+A(-\tau+T)e^{i\phi(T-\tau)}]\nonumber\\
&-&\frac{1}{2}\sin\alpha_3\cos\alpha_4e^{i\theta_3}[e^{i\delta T}A(-\tau-T)e^{i\phi(-\tau-T)}+e^{-i\delta \tau}A(\tau+T)e^{i\phi(\tau+T)}].
\label{eq:Waveform34reduced}
\end{eqnarray}
\end{widetext}
Then we get the Glauber correlation function $G_{P3P4}^{(2)}(T,\tau)=|\psi_{34}(T,\tau)|^2$. With two-photon joint detection efficiency $\eta$, time-bin width $\Delta t$, and total measurement time $\Delta t_m$, the two-photon coincidence counts can be calculated as $C_{P3P4}(T,\tau)=G_{P3P4}^{(2)}(T,\tau)\eta\Delta t \Delta t_m$. To retrieve the phase $\phi(\tau)$, we take the 6 sets of coincidence counts shown in Table \ref{table:tabl1} and have \cite{SupplementalMaterial}
\begin{eqnarray}
\cos[\Lambda(T,\tau)]&=&\frac{[B(T,\tau)+1]}{-2\sqrt{B(T,\tau)}} \frac{C_{\nearrow \nearrow}(T,\tau)-C_{\nearrow \searrow}(T,\tau)}{C_{\nearrow \nearrow}(T,\tau)+C_{\nearrow \searrow}(T,\tau)},\nonumber\\
\label{eq:PhaseDifferenceCos}
\end{eqnarray}
\begin{eqnarray}
\sin[\Lambda(T,\tau)]&=&\frac{[B(T,\tau)+1]}{2\sqrt{B(T,\tau)}} \frac{C_{\nearrow \circlearrowright}(T,\tau)-C_{\nearrow \circlearrowleft}(T,\tau)}{C_{\nearrow \circlearrowright}(T,\tau)+C_{\nearrow \circlearrowleft}(T,\tau)}, \nonumber\\
\label{eq:PhaseDifferenceSin}
\end{eqnarray}
where $B(T,\tau)=C_{\updownarrow\leftrightarrow}(T,\tau)/C_{\leftrightarrow\updownarrow}(T,\tau)$, and $\Lambda(T,\tau)=\Xi(T,\tau)+\Lambda_0$. $\Lambda_0$ is the residual phase constant resulting from imperfections of the optical components. The combined phase difference is defined as $\Xi(T,\tau)=\phi(\tau+T)-\phi(\tau-T)-\delta T$ for $\tau > T$, and $\Xi(T,\tau)=\phi(-\tau-T)-\phi(-\tau+T)+\delta T$ for $\tau < -T$.  With $\phi(\tau)=\phi(-\tau)$, we have $\Xi(T,\tau)=-\Xi(T,-\tau)$. Therefore the residual phase can be determined by $\Lambda_0=\frac{1}{2t_a}\int_{-t_a}^{t_a}\Lambda(T,\tau)]d\tau$. Solving $\Xi(T,\tau)$ and setting a reference phase point $\phi(\tau_0)=0$, we can obtain the phase function by the following recursion
\begin{widetext}
\begin{eqnarray}
\phi(\tau_0+2nT>T)&=&\phi[\tau_0+2(n-1)T]+\Xi[T,\tau_0+(2n-1)T]+\delta T, (n=\pm1, \pm2, ...),\nonumber \\
\phi(-\tau_0-2nT<-T)&=&\phi[-\tau_0-2(n-1)T]+\Xi[T,\tau_0+(2n-1)T]-\delta T, (n=\pm1, \pm2, ...).
\label{eq:PhaseFunction}
\end{eqnarray}
\end{widetext}

\begin{table}
\begin{ruledtabular}
\begin{tabular}{|c|c||l|c|}
  %\hline
  % after \\: \hline or \cline{col1-col2} \cline{col3-col4} ...
  \# & Coincidence counts & $P_3 (\alpha_3, \theta_3)$ & $P_4 (\alpha_4, \theta_4)$ \\ \hline
  1 & $C_{\updownarrow\leftrightarrow}(T,\tau)$ & $\updownarrow (\pi/2,0)$ & $\leftrightarrow$ (0,0)\\ \hline
  2 & $C_{\leftrightarrow\updownarrow}(T,\tau)$ & $\leftrightarrow$ (0,0) & $\updownarrow (\pi/2,0)$\\ \hline
  3 & $C_{\nearrow \nearrow}(T,\tau)$ & $\nearrow (\pi/4,0)$ & $\nearrow (\pi/4,0)$ \\ \hline
  4 & $C_{\nearrow \searrow}(T,\tau)$ & $\nearrow (\pi/4,0)$ & $\searrow (-\pi/4,0)$ \\ \hline
  5 & $C_{\nearrow \circlearrowright}(T,\tau)$ & $\nearrow (\pi/4,0)$ & $\circlearrowright (\pi/4,\pi/2)$ \\ \hline
  6 & $C_{\nearrow \circlearrowleft}(T,\tau)$ & $\nearrow (\pi/4,0)$ & $\circlearrowleft (\pi/4,-\pi/2)$ \\ %\hline
  \end{tabular}
 \end{ruledtabular}
  \caption{\label{table:tabl1} Polarization projection configuration for the two-photon coincidence measurement after the beam splitter. $\updownarrow$: Vertically polarized, $\leftrightarrow$: Horizontally polarized,  $\nearrow$: $45^o$ linearly polarized, $\searrow$: $-45^o$ linearly polarized, $\circlearrowright$: right-circularly polarized, $\circlearrowleft$: left-circularly polarized.}
\end{table}

There is limitation for directly applying Eq.~(\ref{eq:PhaseFunction}) with small $T$ that determines the temporal resolution. For example, if the waveform has some nodes where the amplitude is zero, the phase at the nodes can take any value and thus is totally uncertain. In this case, the phase between amplitude islands are disconnected and their phase difference cannot be resolved when $T$ is too short. Using a long $T$ we can bridge these amplitude islands and determine their phase difference but scarify the time resolution. To solve this problem, we propose a two-step recursion algorism. At the first step, we use a small $T_s$ to obtain a high resolution tomography inside each amplitude island. At the second step, we chose some amplitude peak as the reference point where the phase variation has been precisely determined and use a longer $T_l$ to measure the phase difference between the two islands without touching the node. In this way we can obtain a full phase tomography with a resolution of $2T_s$.

\begin{figure}
\includegraphics[width=\linewidth]{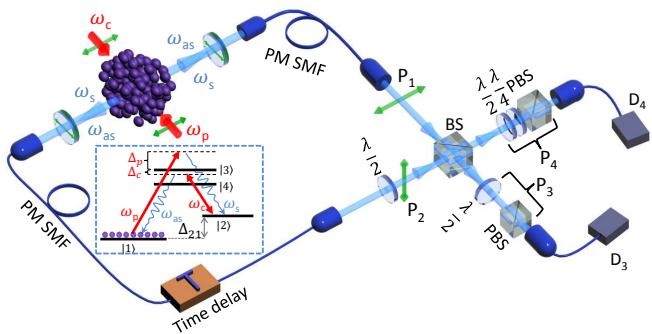}
\caption{\label{fig:ExperimetalSetup} (color online). Experimental setup for biphoton temporal tomography.}
\end{figure}

The experimental setup for our proof-of-principle demonstration is shown in Fig.~\ref{fig:ExperimetalSetup}. We work with a right-angle SFWM with laser cooled $^{85}$Rb atoms in a magneto-optical trap \cite{DuPRA2012}. $|1\rangle=|5S_{1/2}, F=2\rangle$ and $|2\rangle=|5S_{1/2}, F=3\rangle$ are two hyperfine ground states split by $\Delta_{12}=2\pi\times$3036 MHz. Other two relevant atomic energy levels are $|3\rangle=|5P_{3/2}, F=3\rangle$ and $|4\rangle=|5P_{3/2}, F=2\rangle$. The pump laser is far detuned from transition $|1\rangle\rightarrow|3\rangle$ by $\Delta_p$, and the coupling laser is near-resonance to the transition $|2\rangle\leftrightarrow|3\rangle$ (with an adjustable detuning $\Delta_c$).The atomic optic depth is about 3. The phase-matching condition allows spontaneously generating backward paired photons in a right-angle geometry, where the biphotons have two path choices: Stokes photon goes to port 1 and anti-Stokes photon to port 2, or Stokes photon goes to port 2 and anti-Stokes to port 1. In this configuration we do not need the two mirrors M$_1$ and M$_2$ in Fig.~\ref{fig:Schematics}. The linear polarizer P$_3$ comprises of a $\lambda/2$-wave plate and a polarization beam splitter (PBS), and the polarizer P$_4$ comprises of a $\lambda/2$-wave plate, a $\lambda/4$-wave plate and a PBS. The coincidence counts are recorded by D$_3$ and D$_4$ with a time-bin width of 1 ns. For the two-step temporal phase tomography, we set $T_s=$1.0 ns and $T_l=$5.8 ns.

\begin{figure}
\includegraphics[width=\linewidth]{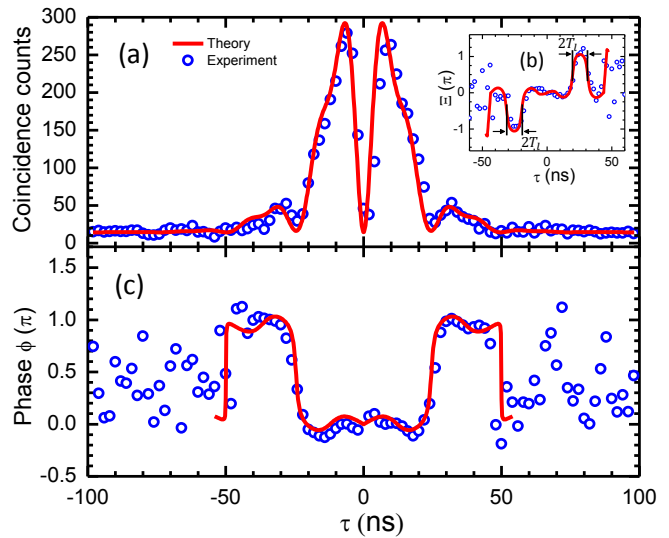}
\caption{\label{fig:Tomography1} (color online). Temporal quantum-state tomography for degenerate biphotons ($\delta$=0) with Rabi oscillation. (a) Two-photon coincidence counts before the BS. (b) The measured combined phase difference function $\Xi(\tau)$ for $T_l$=5.8 ns. (c) The constructed phase $\phi(\tau)$.}
\end{figure}

We start the temporal quantum-state tomography for nearly degenerate biphotons ($\delta\simeq0$), by setting the parameters $\Delta_p=\Delta_{12}-2\pi\times$3 MHz and $\Delta_c=-2\pi\times$10 MHz. The pump laser and coupling laser Rabi frequencies are $\Omega_p=2\pi\times$164.3 MHz and $\Omega_c=2\pi\times$39.2 MHz respectively. The additional pump frequency detuning $\Delta_p-\Delta_{12}=-2\pi\times$3 MHz is to compensate the light shift and residual magnetic-field-induced Zeeman shift. With these parameters, theory predicts the biphoton waveform displays a damped Rabi oscillation with $\pi$-phase flip across the nodes \cite{DuJOSAB2008}. The two-photon coincidence counts $C_{12}(\tau)\propto A^2(\tau)+A^2(-\tau)$ is shown in Fig.~\ref{fig:Tomography1}(a), where the plot at $\tau=t_{as}-t_{s}>0$ is the Stoke to anti-Stokes correlation [$A^2(\tau)$] and that at $\tau=t_{s}-t_{as}<0$ is the anti-Stoke to Stokes correlation [$A^2(-\tau)$]. The solid curve are obtained following the theory in the interaction picture \cite{DuJOSAB2008} that accounts the effect from the atomic energy level $|4\rangle$. The measured combined phase difference function excluding the residual phase constant (for $T_l$=5.8 ns) is shown in Fig.~\ref{fig:Tomography1}(b). Together with the data of $T_s=1.0$ ns, we construct the phase function in Fig.~\ref{fig:Tomography1}(c). The experimental result confirms the theory within the temporal correlation time. The large fluctuation for $|\tau|>50$ ns is caused by the uncertainty of the phase due to the amplitude approaching zero and therefore is reasonable. As shown in Fig.~\ref{fig:Tomography1}(c), the $\pi$-phase jumps across the amplitude nodes are clearly resolved.

\begin{figure}
\includegraphics[width=\linewidth]{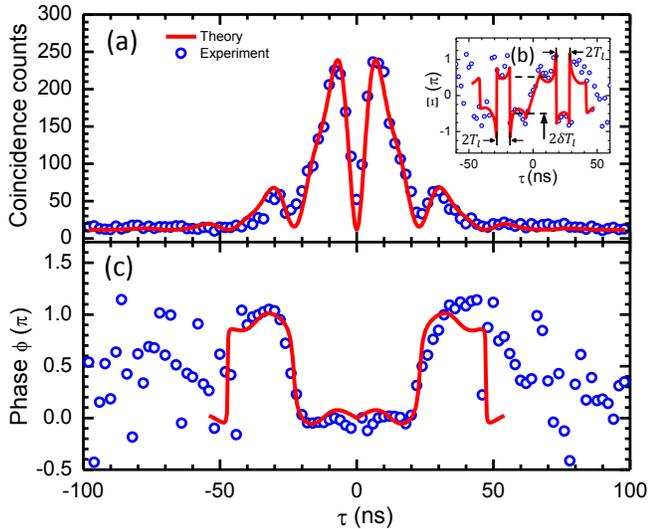}
\caption{\label{fig:Tomography2} (color online). Temporal quantum-state tomography for nondegenerate biphotons ($\delta=2\pi\times$43 MHz) with Rabi oscillation. (a) Two-photon coincidence counts before the BS. (b) The measured combined phase difference function $\Xi(\tau)$ for $T_l$=5.8 ns. (c) The constructed phase $\phi(\tau)$.}
\end{figure}

\begin{figure}
\includegraphics[width=\linewidth]{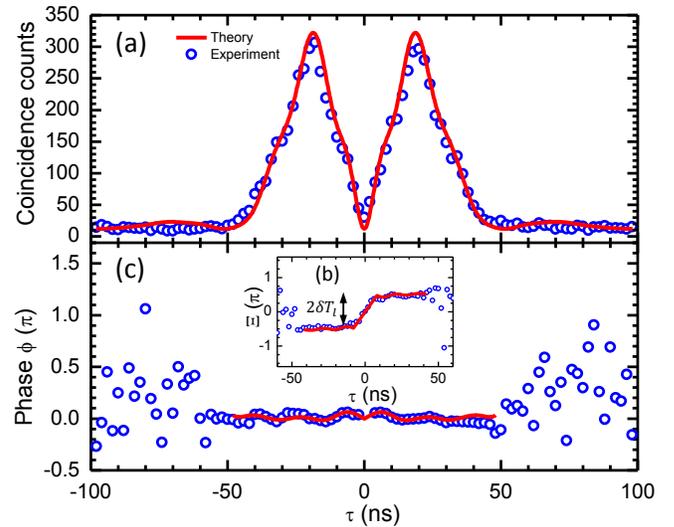}
\caption{\label{fig:Tomography3} (color online). Temporal quantum-state tomography for non-degenerate biphotons ($\delta=2\pi\times$43 MHz) with zero coupling detuning. (a) Two-photon coincidence counts before the BS. (b) The measured combined phase difference function $\Xi(\tau)$ for $T_l$=5.8 ns. (c) The constructed phase $\phi(\tau)$.}
\end{figure}

We next work with nondegenerate biphotons ($\delta=2\pi\times$43 MHz), by setting the parameters $\Delta_p=\Delta_{12}+2\pi\times$40 MHz, $\Delta_c=-2\pi\times$10 MHz, $\Omega_p=2\pi\times$93.3 MHz, and $\Omega_c=2\pi\times$42.6 MHz. The measured $\Xi$ is shown in Fig.~\ref{fig:Tomography3}(b), where the jump across $\tau=0$ is caused by the frequency difference ($2\delta T_l$) as predicted from Eqs.~(\ref{eq:PhaseFunction}). The constructed phase function and theoretical plot are shown in Fig.~\ref{fig:Tomography2}(c).

We then reduce the coupling laser detuning to zero ($\Delta_c=0$) and lower the coupling laser power ($\Omega_c=2\pi\times$19.6 MHz) to prolonger the biphoton correlation time. As shown in Fig.~\ref{fig:Tomography3}(a), the correlation time is extended to about 50 ns and the second oscillation period is completely damped out. The measured combined phase difference function is shown in Fig.~\ref{fig:Tomography3}(b), whose step near $\tau=0$ resolves the frequency difference between the photons. The constructed phase function is shown in Fig.~\ref{fig:Tomography3}(c) and shows the biphoton waveform is Fourier transform limited.

\begin{table}
\begin{ruledtabular}
\begin{tabular}{|c|c|c|}
  %\hline
  % after \\: \hline or \cline{col1-col2} \cline{col3-col4} ...
   & CS$_{max}$ & $g_c^{(2)}$ \\ \hline
  Fig.3 & 204 & 0.27$\pm$0.05\\ \hline
  Fig.4& 89 & 0.16$\pm$0.07\\ \hline
  Fig.5 & 203 & 0.19$\pm$0.02\\ %\hline
  \end{tabular}
 \end{ruledtabular}
  \caption{\label{table:tabl2} Violation of Cauchy-Schrotwz inequality and the measured conditional $g_c^{(2)}$.}
\end{table}

We further measure the nonclassical properties of the photon source by verifying its violation of the Cauchy-Schwartz inequality \cite{ClauserPRD1974}. With normalized cross-correlation function $g_{s,as}^{(2)}(\tau)$ and auto-correlation functions $g_{s,s}^{(2)}(\tau)$ and $g_{as,as}^{(2)}(\tau)$, a classical source follows $CS=[g_{s,as}^{(2)}(\tau)]^2/[g_{s,s}^{(2)}(0)g_{as,as}^{(2)}(0)]\leq1$. Another measure is to confirm the quantum nature of heralded single photons with the conditional autocorrelation function $g_c^{(2)}$ intergraded over entire waveform temporal length. The masured CS and intergrated $g_c^{(2)}$ of our biphoton source are shown in Table \ref{table:tabl2}. In all the cases, the Cauchy-Schwartz inequality is violated, and $g_c^{(2)}$ is well below the two-photon threshold of 0.5.

In conclusion, we proposed and demonstrated a temporal quantum-state tomography for narrowband biphotons generated from SFWM in cold atoms. In all three degenerate and non-degenerate cases, experiments agree perfectly with the theory. The degeneracy of the photon frequencies does not need to be preassumed and the frequency difference can be determined from the measured $\Xi(T,\tau)$. Although our proof-of-principle demonstration takes the advantages of the right-angle configuration, the tomography method can be applied to a general case following the setup in Fig.~\ref{fig:Schematics}.

The work was supported by the Hong Kong Research Grants Council (Project No. 601113).

%%%%%%%%%%%%%%%%  Reference  %%%%%%%%%%%%%%%%%%%%%%%%%%%%


\begin{thebibliography}{99}

\bibitem{KwiatPRL1999} A. G. White, D. F. V. James, P. H. Eberhard, and P. G. Kwiat, Phys. Rev. Lett. \textbf{83}, 3103 (1999).

\bibitem{WhitePRA2001} D. F. V. James, P. G. Kwiat, W. J. Munro, and A. G. White, Phys. Rev. A \textbf{64}, 052312 (2001).

\bibitem{AdamsonPRL2010} R. B. A. Adamson and A. M. Steinberg, Phys. Rev. Lett. \textbf{105}, 030406 (2010).

\bibitem{LvovskyPRL2001} A. I. Lvovsky, H. Hansen, T. Aichele, O. Benson, J. Mlynek, and S. Schiller, Phys. Rev. Lett. \textbf{87}, 050402 (2001).

\bibitem{OurjoumtsevPRL2006} A. Ourjoumtsev, R. Tualle-Brouri, and P. Grangier, Phys. Rev. Lett. \textbf{96}, 213601 (2006).

\bibitem{MacRaePRL2012} A. MacRae, T. Brannan, R. Achal, and A. I. Lvovsky, Phys. Rev. Lett. \textbf{109}, 033601 (2012).

\bibitem{LvovskyRMP2009} A. I. Lvovsky and M. G. Raymer, Rev. Mod. Phys. \textbf{81}, 299 (2009).

\bibitem{LauratPRL2013} O. Morin, C. Fabre, and J. Laurat, Phys. Rev. Lett. \textbf{111}, 213602 (2013)

\bibitem{Lvovsky2014} Z. Qin, A. S. Prasad, T. Brannan, A. MacRae, A. Lezama, and A. I. Lvovsky, arXiv:1405.6251 [quant-ph].

\bibitem{OuPRL1992} Z. Y. Ou, S. F. Pereira, H. J. Kimble, and K. C. Peng, Phys. Rev. Lett. \textbf{68}, 3663 (1992).

\bibitem{SchoriPRA2002} C. Schori, J. L. S{\o}rensen, and E. S. Polzik, Phys. Rev. A \textbf{66}, 033802 (2002).

\bibitem{LauratJOB2005} J. Laurat, G. Keller, J. A. Oliveira-Huguenin, C. Fabre, T. Coudreau, A. Serafini, G. Adesso, and F. Illuminati, J. Opt. B: Quantum Semiclassical Opt. \textbf{7}, S577 (2005).

\bibitem{VasilyevPRL2000} M. Vasilyev, S.-K. Choi, P. Kumar, and G Mauro DAriano, Phys. Rev. Lett. \textbf{84}, 2354 (2000).

\bibitem{DAuriaPRL2009} V. D'Auria, S. Fornaro, A. Porzio, S. Solimeno, S. Olivares, and M. G. A. Paris, Phys. Rev. Lett. \textbf{102}, 020502 (2009).

\bibitem{PanPRL2008} X. H. Bao, Y. Qian, J. Yang, H. Zhang, Z.-B. Chen, T. Yang, and J.-W. Pan, Phys. Rev. Lett. \textbf{101}, 190501 (2008).

\bibitem{ScholzPRL2009} M. Scholz, L. Koch, and O. Benson, Phys. Rev. Lett. \textbf{102}, 063603 (2009).

\bibitem{FeketePRL2013} J. Fekete, D. Rielander, M. Cristiani, and H. de Riedmatten, Phys. Rev. Lett. \textbf{110}, 220502 (2013).

\bibitem{Subnatural} S. Du, P. Kolchin, C. Belthangady, G. Y. Yin, and S. E. Harris, Phys. Rev. Lett. \textbf{100}, 183603 (2008).

\bibitem{YanPRL2014} K. Liao, H. Yan, J. He, S. Du, Z.-M. Zhang, S.-L. Zhu, Phys. Rev. Lett. \textbf{112}, 243602 (2014).

\bibitem{DuOptica2014} L. Zhao, X. Guo, C. Liu, Y. Sun, M. M. T. Loy, and S. Du, Optica \textbf{1}, 84 (2014).

\bibitem{EOMSinglephoton} P. Kolchin, C. Belthangady, S. Du, G. Y. Yin, and S. E. Harris, Phys. Rev. Lett. \textbf{101}, 103601 (2008).

\bibitem{DuPRA2009} S. Du, J. Wen, and C. Belthangady, Phys. Rev. A \textbf{79}, 043811 (2009).

\bibitem{DuPRL2010Biphoton} J. F. Chen, S. Zhang, H. Yan, M. M. T. Loy, G. K. L. Wong, and S. Du, Phys. Rev. Lett. \textbf{104}, 183604 (2010).

\bibitem{SinglePhotonPrecursor} S. Zhang, J. F. Chen, C. Liu, M. M. T. Loy, G. K. L. Wong, and S. Du, Phys. Rev. Lett. \textbf{106}, 243602 (2011).

\bibitem{DuOptExpress2012} S. Zhou, S. Zhang, C. Liu, J. F. Chen, J. Wen, M. M. T. Loy, G. K. L. Wong, and S. Du, Opt. Express \textbf{20}, 24124 (2012).

\bibitem{DuPRL2012} S. Zhang, C. Liu, S. Zhou, C.-S. Chuu, M. M. T. Loy, and S. Du, Phys. Rev. Lett. \textbf{109}, 263601 (2012).

\bibitem{ClauserPRD1974} J. F. Clauser, Phys. Rev. D \textbf{9}, 853 (1974).

\bibitem{GrangierEPL1986} P. Grangier, G. Roger, and A. Aspect, Europhys. Lett. \textbf{1}, 173 (1986).

\bibitem{DuJOSAB2008} S. Du, J. Wen, and M. H. Rubin, J. Opt. Soc. Am. B \textbf{25}, C98 (2008).

\bibitem{SupplementalMaterial} See the Supplemental Material for a detailed derivation.

\bibitem{DuPRA2012} C. Liu, J. F. Chen, S. Zhang, S. Zhou, Y.-H. Kim, M. M. T. Loy, G. K. L. Wong, and S. Du, Phys. Rev. A \textbf{85}, 021803(R) (2012).

\end{thebibliography}
\end{document}